\documentclass[twocolumn,reprint,aps,superscriptaddress,nofootinbib]{revtex4-2}

\usepackage{amsmath, amssymb, amsfonts, graphics, setspace}

\usepackage{enumitem}

\usepackage{graphicx}
\usepackage{lmodern}
\usepackage[T1]{fontenc}
\usepackage{txfonts}
\usepackage{color}
\usepackage{xcolor}

\newcommand{\nn}{\nonumber}

\tabularnewline

\def \be {\begin{equation}}
\def \ee {\end{equation}}
\def \bea {\begin{eqnarray}}
\def \eea {\end{eqnarray}}
\newcommand{\eq}[1]{(\ref{#1})}
\newcommand{\rd}{\color[rgb]{0,0,0}}

\begin{document}
\title{Weak Cosmic Censorship and Second Law of Black Hole Thermodynamics \\ {\rd in Higher Derivative Gravity} }

\author{Feng-Li Lin}
\affiliation{Department of Physics, National Taiwan Normal University, Taipei, Taiwan 116}
\author{Bo Ning}
\affiliation{College of Physics, Sichuan University, Chengdu, Sichuan 610064, China}
\affiliation{Peng Huanwu Center for Fundamental Theory, Hefei, Anhui 230026, China}
\author{Yanbei Chen}
\affiliation{Burke Institute of Theoretical Physics and Theoretical Astrophysics 350-17, California Institute of Technology, Pasadena, California 91125}

\date{\today}
\begin{abstract}
Infalling matter may destroy a black hole and expose the naked singularity. Thus, Penrose proposed the weak cosmic censorship conjecture to avoid such a possibility. On the other hand, if the black hole is not destroyed by infalling matter, from the second law of black hole thermodynamics, the black hole entropy should increase due to the information carried by the infalling matter. In this work, we demonstrate by examples of perturbative near-extremal black holes in higher derivative gravity theories that the second law implies weak cosmic censorship. We also compare our proposal to the one developed by Sorce and Wald based on the first law of black hole thermodynamics and show that the latter fails to yield weak cosmic censorship in such cases. Finally, we give proof of our proposal for generic gravity theories. 
\end{abstract}

\maketitle

\section{Introduction} 
Black holes are the simplest objects predicted by general relativity --- with intriguing features. 
Even though black holes have curvature singularities, around which tidal gravity diverges and physical laws break down, in analytic black hole solutions, these singularities are always shielded by the event horizon~\cite{Penrose:1964wq}.  Penrose further proposed \cite{Penrose:1969pc} the weak cosmic censorship conjecture (WCCC): the curvature singularity will always be hidden behind the horizon for {\it generic} black holes, i.e., no naked singularities.  Another intriguing feature is that the first and second laws of thermodynamics govern black holes.  Bekenstein's conjecture that a black hole's entropy must be proportional to its area~\cite{Bekenstein:1973ur,Bekenstein:1974ax} was substantiated by the theoretical discovery of Hawking Radiation, and the fact that this {\it thermal radiation} has a temperature proportional to the black hole's surface gravity~\cite{Hawking:1975vcx}.

Naively, one shall expect the connection between WCCC and the second law. As the second law requires, the entropy of a black hole can never decrease. This prevents the appearance of a naked singularity. The proof for the second law for Einstein gravity given in \cite{Christodoulou:1970wf, Hawking:1971tu, Hawking:1971vc} can imply WCCC, later more direct connection is discussed in \cite{Sorce:2017dst}. However, a demonstration for modified gravities is nontrivial, since, in this case, the entropy follows Wald's entropy formula \cite{Wald:1993nt,Iyer:1994ys} but not the area law. Thus, the second law may not ensure the area increase, and the above connection is unclear.  To ensure WCCC is a universal physical principle, in this paper, we demonstrate such a connection explicitly for modified gravities by showing that the WCCC follows as long as the second law holds.

Wald started the demonstration of WCCC by gedanken experiments that attempt to destroy the horizon by overcharging or overspinning a black hole with infalling matter~\cite{Wald:I,Sorce:2017dst}. 
For simplicity, we shall focus on overcharging non-spinning black holes.  Assume a family of electro-vacuum {\it solutions} to the gravitational and electromagnetic field equations, parametrized by mass $m$ and charge $q$ to describe the configurations before and after the matter  ``falls in''. We denote the condition for the space-time to be a black hole, i.e., with a horizon that covers the singularity, by 
\be\label{WCCC_c}
W(m,q)\ge 0. 
\ee
The exact form of $W(m,q)$ depends on the underlying theory. For example, the (outer) horizon of a Reissner–Nordström black hole is $r_+=m+ \sqrt{m^2-q^2}$, thus $W(m,q)=m^2-q^2$ so that \eq{WCCC_c} guarantees a positive and real $r_+$, thus the existence of a  horizon.

The demonstration of WCCC is to show $W(m+\Delta m, q+\Delta q)\ge 0$ given the initial mass $m$ and charge $q$, for all the respective {\it physically allowed} changes $\Delta m$ and $\Delta q$ due to the infalling matter. Values of $\Delta m$ and $\Delta q$ depend on how the matter falls in and the underlying gravity theory. Intuitively, the physical constraints on $\Delta m$ and $\Delta q$ should come from the laws of black hole dynamics. Indeed, in \cite{Chen:2020hjm}, we have demonstrated that the first law of black hole dynamics is a universal condition to guarantee WCCC for extremal black holes in generic gravity theories.  For near-extremal black holes, Sorce and Wald \cite{Sorce:2017dst} generalized the first law constraint to second-order variations and showed that this could guarantee WCCC in Einstein-Maxwell theory \footnote{Christodoulou proved that naked singularity can occur in Einstein-scalar system though is unstable, hence the cosmic censorship is still preserved \cite{Christodoulou:1994}. }.

One will face some challenges when trying to generalize the approach of \cite{Sorce:2017dst} to modified gravities. The main challenge is to unambiguously define the canonical energy of gravitational waves for modified gravity and the respective energy condition required to ensure WCCC.  Without such energy conditions, one can only consider the spherical collapsing with no induced gravitational wave.  Moreover, in \cite{Sorce:2017dst}, the canonical energy is evaluated by relating it to the black hole entropy by the first law. Still, such a substitution is unclear due to the ambiguity of canonical energy in modified gravity.

Indeed, we show that the approach of \cite{Sorce:2017dst} fails to demonstrate WCCC for the modified gravities. To bypass the aforementioned challenges and remedy the resultant failure, we propose demonstrating the WCCC with the second law. Our proposal does not need canonical energy or conditions for matter and gravity. All we need is Wald's formula for black hole entropy. In Einstein's gravity, the first law and energy condition can guarantee the second law, but it is unclear for modified gravities. The result obtained here may also shed some light on this issue. Finally, before we proceed, we shall emphasize the demonstration is not a tautology. 
Although the existence of entropy is the premise of the second law, itself does not guarantee the WCCC condition \eq{WCCC_c}, since a decreasing entropy would indicate naked singularity in general relativity according to \cite{Sorce:2017dst}. Thus, our demonstration is a consistency check in the same spirit of \cite{Sorce:2017dst}.

\section{WCCC condition in higher derivative gravity theories}
We consider the general quartic order corrections to Einstein-Maxwell theory, which is given by the following Lagrangian:
\bea\label{eq:action}
L  &=& \frac{1}{2\kappa}R -\frac{1}{4}F_{\mu\nu}F^{\mu\nu}+ c_1 R^2+ c_2 R_{\mu\nu}R^{\mu\nu}+c_3 R_{\mu\nu\rho\sigma}R^{\mu\nu\rho\sigma}\;  \nn \\ \nn
  &&+ c_4 \kappa R F_{\mu\nu}F^{\mu\nu}+ c_5 \kappa R_{\mu\nu}F^{\mu\rho}F^{\nu}{}_{\rho} +
       c_6 \kappa R_{\mu\nu\rho\sigma}F^{\mu\nu}F^{\rho\sigma} \;\\   
  &&+ c_7 \kappa^2 F_{\mu\nu}F^{\mu\nu}F_{\rho\sigma}F^{\rho\sigma} + c_8 \kappa^2
        F_{\mu\nu}F^{\nu\rho}F_{\rho\sigma}F^{\sigma\mu} 
\eea
where $\kappa=8\pi G_N$, which will be set to $2$ below, and $c_i$'s are dimensionless constants. From the point of view of effective field theory, the above higher derivative theories can arise naturally from quantum corrections. Thus, some of these theories can be the genuine description of low energy black hole dynamics but remains experimentally elusive due to smallness of  $c_i$'s. If WCCC is a fundamental principle for protecting the predictive power of theory, it should also apply to generic effective field theories of gravity.

To study WCCC, we first generalize the perturbative method of \cite{Kats:2006xp} to solve the charged black hole solutions up to ${\cal O}(c_ic_j)$ with $i,j=1,\cdots 8$ \footnote{The detailed solutions can be found in Appendix {\ref{sec:1}}.}. Based on these solutions, we can find the following $W(m,q)$ for \eq{WCCC_c},
\be\label{WCCC_c2c4}
W(m, q)=m^2-q^2\Bigg(1-\frac{4 c_0}{5 q^2}+\frac{128 c_4^2}{21 q^4 }+ \cdots \Bigg)^2 
\ee
with $c_0 \equiv  c_2 + 4c_3 + c_5 + c_6 + 4c_7 + 2c_8$, and $\cdots$ denotes the other ${\cal O}(c_ic_j)$ terms. For simplicity, below we will only show the result for the case with nonzero $c_4$ as a demonstration. The other cases with nonzero $c_{i\ne 4}$ can be found in the supplementary material. Besides, the black hole entropy can be obtained  by Wald's formula \cite{Wald:1993nt,Bhattacharyya:2021jhr}, and it yields
\bea
\label{eq:SWald}
S(m, q) &=& -2\pi A_h \; \Big[ -\frac{1}{2} -4 c_1 R -4 c_2 R^{rv} + 8 c_3 R^{rvrv} \nn \\
&& + 4 \; (2 c_4 + c_5 + 2 c_6)  F^{rv}F^{rv} \Big]\;.
\eea
where the area of the horizon $A_h$, the curvatures, and field strengths are evaluated by the on-shell solution.

\section{Check WCCC in Wald's gedanken experiment by second law constraints}
\begin{figure} 
  \centering
  \includegraphics[width=0.75\linewidth]{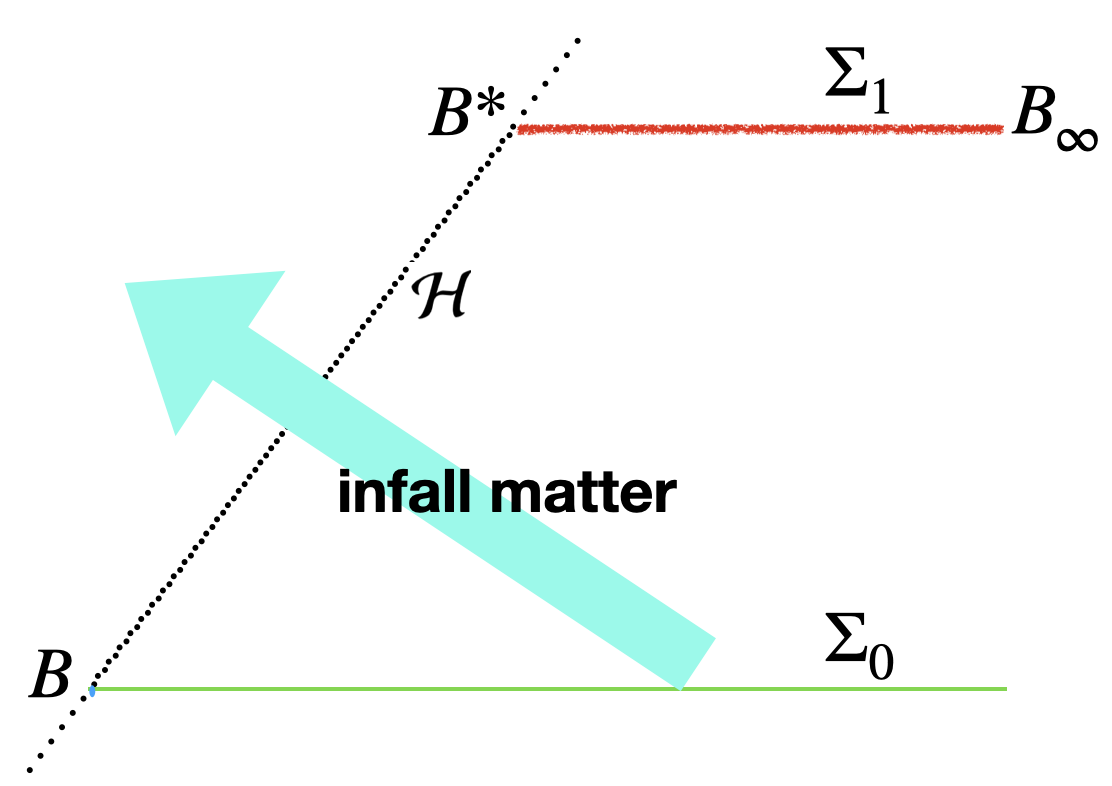}
  \caption{Wald's gedanken experiment by throwing the charged matter into a black hole. The infalling matter crosses the horizon $\cal H$ within a finite time interval.}
  \label{fig:gedanken-plot}
\end{figure}

To follow Wald's gedanken experiment, we consider charged matter falling through the black hole's horizon within a finite time interval. Then, the black hole and the infalling matter are settled to a final stationary state belonging to the same family of solutions, either a new black hole or a naked singularity. The scheme is shown in Fig. \ref{fig:gedanken-plot}.
 
As argued by Sorce and Wald \cite{Sorce:2017dst}, for a near-extremal black hole WCCC might be violated from first-order considerations \cite{Hubeny:1998ga}, but in fact is preserved at second order. Therefore, we need to consider the variations of $m$ and $q$ caused by the infalling matter up to the second order. Here we outline the steps of checking WCCC upon the second law of black hole (thermo)dynamics, which basically require that the  entropy difference between $B$ and $B^*$ due to infalling matter through ${\cal H}$ (see Fig. \ref{fig:gedanken-plot}) is non-decreasing.

Let us consider an initial black hole with $(m,q)$, with a one-parameter family of infalling matter, finally settling down to a new solution with 
 \begin{equation}
 m(\lambda)=m + \lambda \delta m + \frac{\lambda^2 \delta^2 m }{2} \,,\quad
 q(\lambda)=q + \lambda \delta q + \frac{\lambda^2 \delta^2 q}{2} \,.
\end{equation}
Here we keep mass and charge increases up to second order in $\lambda$.  We shall also restrict ourselves to nearly extremal black holes, and for the moment only consider $c_4$.  The initial black hole is second-order away from being extremal: 
    \be 
    q=\sqrt{1-\epsilon^2}\Big(m - \frac{128 c_4^2}{21 m^3}\Big)\;.
    \ee 
Similar to Sorce and Wald, we shall assume $\epsilon$ and $\lambda$ to be of the same order of smallness, and check whether $W(m,q)>0$ is satisfied up to second order. More specifically, we need to check whether constraints on $(\delta m, \delta q,\delta^2 m,\delta^2 q)$, arising from 
\begin{equation}
    S(m(\lambda),q(\lambda)) \ge S(m,q)
\end{equation}
will guarantee $W(m+\Delta m, q+\Delta q)\ge 0$. 

Since we will consider up to second-order variations, we assume the first order variation due to the infalling matter to be optimally done with the second law being satisfied marginally:
    \be\label{dS_1}
    \delta S= \frac{\partial S}{\partial m} \delta m + \frac{\partial S}{\partial q} \delta q=0\;.
    \ee
    Solving this condition gives a relation between $\delta m$ and $\delta q$. For the $c_4$ case that we show explicitly, it yields \be\label{dS_11}
    \delta m = \Bigg[1-\epsilon - \frac{64(2+1098 \epsilon) c_4^2}{7 m^4} \Bigg] \delta q +{\cal O}(\epsilon^2) \;.
    \ee 
    
    For extremal black holes, we have $\epsilon=0$, up to ${\cal O}(c_i)$, we can truncate the terms of ${\cal O}(c_i^2)$ in the above to show that the first law $\delta S\ge 0$ gives 
    \be 
    \delta m \ge \Big(1+\frac{4 c_0}{5 q^2}\Big) \delta q.
    \ee
    This is just the WCCC condition for the extremal black holes, as demonstrated in \cite{Chen:2020hjm} via Sorce-Wald.
    
Let us now consider second order variations due to the infalling matter such that the second law holds, i.e., 
\bea
    \delta^2 S &=& \frac{\partial^2 S}{\partial m^2} (\delta m)^2 + 2 \frac{\partial^2 S}{\partial m \partial q} \delta m \delta q+ \frac{\partial^2 S}{\partial q^2} (\delta q)^2 \nn \\
    &&+ \frac{\partial S}{\partial m} \delta^2 m + \frac{\partial S}{\partial q} \delta^2 q \ge 0\;. \label{dS_2}
\eea 
    For the $c_4$ case, combining this equation with \eq{dS_11}, we obtain
\bea 
    \delta^2 m &\ge & \Bigg[\frac{1-\epsilon}{m} + \frac{256(1655-17372 \epsilon+33099 \epsilon^2)c_4^2}{21 m^5} \Bigg] (\delta q)^2 \nn \\
    &&+ \Bigg[1-\epsilon + \frac{\epsilon^2}{2} -\frac{64(2+1098 \epsilon -8815 \epsilon^2 )c_4^2}{7 m^4}  \Bigg]\delta^2 q. \qquad \; \label{dS_22}
\eea 
This leads to
\be
W(\lambda)=\Bigg(\epsilon\Big(\frac{256 c_4^2}{21 m^3}-m \Big)+ \lambda \Big(1+ \frac{211072 c_4^2}{21 m^4} \Big) \delta q \Bigg)^2 + {\cal O}(c_4^3, \epsilon^3, \lambda^3)
\ee
where the ${\cal O}(c_4^3, \epsilon^3, \lambda^3)$ denotes the higher order terms which will be omitted later for simplicity. Thus, we can conclude that WCCC is preserved by the second law constraints up to ${\cal O}(c_4^2)$. If we consider $W(\lambda)$ only up to ${\cal O}(c_i)$, it takes a simple but not positive definite form \bea
W(\lambda)&=&(\epsilon m - \lambda \delta q)^2+  \frac{8}{5m^2} (\epsilon m - \lambda \delta q)  \nn\\
  && \times \Big(c_0(\epsilon m + 3\lambda \delta q) +10 c_6 \lambda \delta q \Big)+ {\cal O}(c_i c_j)\;.  \label{WCCC_key}
\eea
Completing the square of \eq{WCCC_key} requires ${\cal O}(c_i c_j)$ terms. This is why we need to use the near-extremal black hole solutions up to ${\cal O}(c_i c_j)$ to check WCCC. This is in the same spirit as invoking second-order variations in \cite{Sorce:2017dst} to remedy the earlier mistake of \cite{Hubeny:1998ga} in checking WCCC.
 The check of WCCC for the other cases with $c_{i\ne 4}$ and the case of $c_2$ and $c_4$ can be found in Appendix {\ref{sec:1}}.  All results are consistent with our proposal that the second law constraints imply WCCC \footnote{Especially, the Einstein-Maxwell-Gauss-Bonnet theory with $c_1=c_3=-{1\over 4}c_2$ gives no contribution to the ${\cal O}(c_i)$ term of \eq{WCCC_key}, thus preserving the WCCC. We demonstrate this by the spherical thin-shell collapse in Appendix {\ref{sec:3}}.}.
 
Finally, due to the complication of solving the rotating charged black holes with $c_i$ corrections, we do not check the WCCC for such cases. However, it is straightforward to check WCCC by our second law formalism for Kerr-Newman black holes with its spin denoted by $j$ in Einstein-Maxwell theory and the result is
\be \label{Kerr_Newman}
W(\lambda)=\Bigg( \frac{(j^2-m^4) q \delta q -2 j m^2 \delta j}{m(m^4 + j^2)} \lambda + m\epsilon \Bigg)^2 + {\cal O}(\epsilon^3, \lambda^3) \;.
\ee
The form of \eq{Kerr_Newman} is exactly the same as the one in \cite{Sorce:2017dst}. This shows that our second law proposal yields the same WCCC result as ensured by the first law one of Sorce-Wald.

\section{Comparison with Sorce-Wald formalism}
For comparison, we will show that Sorce-Wald formalism fails to yield WCCC for the modified gravities considered above. Sorce-Wald formalism \cite{Sorce:2017dst} uses the first law constraints to check WCCC. At the second-order variations, one needs to take into account the energy contribution from the induced gravitational and electromagnetic waves, which makes the problem technically involved. For simplicity, we assume the infalling matter is spherical symmetric so that no such waves will be induced.


Our steps outlined earlier to verify WCCC are inspired by the Sorce-Wald formalism. The only difference is that we shall replace the second law constraints by the first law ones. The latters take the following general form \cite{Hollands:2012sf,Sorce:2017dst}
\bea
&& \delta^n m_{\rm ADM} -\Phi_{\rm H} (\delta^n q_{\rm H} + \delta^n q_B) - T_H \delta^n S_B  \nn \\
&& = \delta_{n,2} {\cal E}_{\Sigma}(\phi; \delta \phi, {\cal L}_{\xi}\phi)-\int_{\cal H} \xi^a \epsilon_{ebcd} \delta^n T^{~e}_a \ge \delta_{n,2} {\cal E}_{\Sigma}(\phi; \delta \phi, {\cal L}_{\xi}\phi)\;. \qquad \label{1st_law_gen}
\eea
Here $n=1,2$ is the order of variation, $\xi^{\mu}$ is the timelike Killing vector of the background metric, and $\Phi_H=-\xi^{\mu} A_{\mu}|_{r=r_+}$ is the chemical potential on the horizon. We have used the energy condition on the stress tensor $\delta^n T_{ab}$ of the infalling matter to arrive the last inequality. Sorce-Wald assumed no matter around the bifurcation sphere $B$ of Fig. \ref{fig:gedanken-plot} so that the variations of charge and Wald's entropy vanish, i.e., $\delta^n q_B=\delta^n S_B=0$. On the other hand, when considering the standard first law without source perturbation, we will instead set $\delta^n q_{\rm H}=\delta^n T_{ab}=0$ \footnote{In Appendix {\ref{sec:2}} we check the  first law for $n=1$ for the black holes considered in this work.}.  The higher derivative corrections to Einstein-Maxwell theory cannot affect the ADM mass $m_{\rm ADM}$ and the charge $q_{\rm H}$ of the black hole  due to their higher powers of  $1/r$ suppression.  This implies $\delta^n m_{\rm ADM}=\delta^n m$ and $\delta^n q_{\rm H}=\delta^n q$ as in Einstein-Maxwell theory. The gravitational energy ${\cal E}_{\Sigma}$ on the Cauchy surface $\Sigma={\cal H} \cup \Sigma_1$ of Fig. \ref{fig:gedanken-plot} is the self-gravitating effect, thus is absent for $n=1$. Moreover, since no wave is induced around $\cal H$, ${\cal E}_{\Sigma}={\cal E}_{\Sigma_1}$.

Assume the first law constraint of $n=1$ is optimally done, i.e., $\delta m - \Phi_H \delta q=0$, for the $c_4$ case, it explicitly gives 
\be\label{1st_SW_11}
    \delta m = \Bigg[1-\epsilon - \frac{64(2 - 22\epsilon ) c_4^2}{7 m^4} \Bigg] \delta q  +{\cal O}(\epsilon^2)
\ee 
which is different from \eq{dS_11} at ${\cal O}(\epsilon  c_4^2)$.  To evaluate ${\cal E}_{\Sigma_1}$ when considering the $n=2$ case, Sorce and Wald assumed that the late-time perturbation $\delta \phi$ approaches a stable linear on-shell configuration $\delta \phi^{\rm linear}$, and one can apply \eq{1st_law_gen} of $n=2$ on $\Sigma_1$ with $\delta^2 m=\delta^2 q=0$ so that 
\be\label{Canon_E}
{\cal E}_{\Sigma_1}(\phi; \delta \phi, {\cal L}_{\xi}\phi)={\cal E}_{\Sigma_1}(\phi; \delta \phi^{\rm linear}, {\cal L}_{\xi}\phi)=-T_H \delta^2 S^*
\ee 
where $T_H$ is the Hawking temperature \footnote{Due to our convention for $m$ and $q$ by a scale factor $1/4\pi$, here $T_H$ is the scaled Hawking temperature by the same factor.} of the initial black hole, but the variation of Wald's entropy $\delta^2 S^*$ is evaluated at $B^*$ of Fig. \ref{fig:gedanken-plot} with respect to $\phi+\delta \phi^{\rm linear}$. By construction $\delta^2 S^*=\delta^2 S^*(\delta m,\delta q)$\,, 
{\rd hence} the $n=2$ first law constraint now takes a second-law-like form
\be\label{2nd_SW_f} 
\delta^2 S^*(\delta m, \delta q) + \frac{1}{T_H} \Big( \delta^2 m -\Phi_H \delta^2 q \Big) \ge 0\;.
\ee
With the help of \eq{1st_SW_11}, for the case $c_4$ we show explicitly,  \eq{2nd_SW_f} gives
\bea 
 \delta^2 m &\ge & \Bigg[\frac{1-\epsilon}{m} - \frac{256(1285-9088 \epsilon+33261 \epsilon^2)c_4^2}{21 m^5} \Bigg] (\delta q)^2 \nn \\
    &&+ \Bigg[1-\epsilon + \frac{\epsilon^2}{2} -\frac{64(2-22 \epsilon +145 \epsilon^2 )c_4^2}{7 m^4}  \Bigg]\delta^2 q \;, \qquad \label{2nd_SW_22}
\eea 
which is different from \eq{dS_22} at ${\cal O}(c_4^2)$.  
Based on \eq{1st_SW_11} and \eq{2nd_SW_22}, we can evaluate $W(m(\lambda),q(\lambda))$ for the case $c_4$ and the result is 
\be
W(\lambda)=\Bigg(\epsilon\Big(\frac{161024 c_4^2}{21 m^3}+ m \Big)- \lambda \Big(1- \frac{165248 c_4^2}{21 m^4} \Big) \delta q \Bigg)^2 - \frac{15360 \epsilon^2 c_4^2}{m^2}
\ee
which cannot be completed the square at ${\cal O}(c_4^2)$ to protect WCCC. Similar results for the others cases of $c_{i\ne 4}$ and of $c_2$ and $c_4$ up to ${\cal O}(c_ic_j)$ can be found in Appendix {\ref{sec:1}} . 

To conclude our work, in the following we outline a general proof of WCCC based on our second law proposal. 

\section{Proof of WCCC in general}
Suppose we have $m$ and $q_j$, and $m = m_{\rm ex}(q_j)$ is the mass of extremal black holes, with black holes given by $m\ge m_{\rm ex}(q_j)$.   Let us define $\mu = m - m_{\rm ex}(q_j)$, which encodes the (deviation from the) extremality condition.  Let us also denote $r_h(\mu,q_j)$ the horizon radius, with $R(q_j) = r_h(0,q_j)$ the radius of extremal black holes as a function of $q_j$.  In Appendix {\ref{sec:4}}, we argue that for $\mu$ inside an open neighborhood of 0, 
\begin{equation}
\label{eqrexp}
r_h(\mu,q_j ) = R(q_j) + \sqrt{\mu} \,\rho (q_j,\sqrt{\mu})
\end{equation}
with $\rho$ a smooth function of its two arguments. Suppose a quantity like the Wald entropy can still be defined in a modified theory of gravity, and that it is expressed as a smooth function of $r_h$, $m$ and $q_j$. Of course, we can also express it in terms of $m$ and $q_j$, but that expression may not be infinitely smooth in an open neighborhood of the extremal boundary.   Let us write
\begin{equation}
S = S(r_h,\mu,q_j)
\end{equation}
with $\partial S/\partial r_h \neq   0$. 
For example, the $S$ defined in \eq{eq:SWald}, is of this form. Since $A=4\pi r_h^2$, and the correction terms are expected to be much less than unity, $\partial S/\partial r_h$ is  non-zero.  For a family of solutions parametrized by $\lambda$, we require that $S(\lambda>0) \ge S(\lambda=0)$ still holds, as a generalized second-law of black-hole thermodynamics.  

Let us now start from a configuration with $(\mu,q_j) = (\epsilon^2,q_{j0})$, with $\epsilon>0$ a small quantity, and deviate away from it with 
\bea
\label{muexp}
\mu&=&\epsilon^2 + \delta\mu \lambda+\delta^2\mu \frac{\lambda^2}{2}\;, \\
q_j &=& q_{j0} +\delta q_{j}\lambda+ \delta^2q_{j}\frac{\lambda^2}{2}\;.
\eea
Note that the deviation from the extremality is ${\cal O}(\epsilon^2)$. We will treat $\epsilon$ and $\lambda$ as quantities with the same order of smallness, and  use the fact that $dS/d\lambda $ and $d^2S/d\lambda^2$ should be finite at $\lambda =0$, as $\epsilon\rightarrow 0$.  
For $dS/d\lambda $, we have {\rd a leading contribution of }
\begin{equation}\label{d1S_g}
\frac{dS}{d\lambda}\bigg|_{\lambda =0} {\rd \,\sim\,} \;
\frac{\partial S}{\partial r_h} \frac{\rd \rho}{2\epsilon}\delta\mu  
\end{equation} 
where we have used $\mu = \epsilon^2$ for $\lambda =0$.  Here in order for $dS/d\lambda $ to be finite, we will require  $\delta\mu\sim \epsilon$.  Inserting this into the second derivative, we obtain 
\begin{equation}\label{d2S_g}
\frac{d^2S}{d\lambda^2}\bigg|_{\lambda =0} 
{\rd \,\sim\,} \; \frac{\rd \rho}{2\epsilon^3}\left(\epsilon^2 \delta^2\mu-\frac{1}{2}\delta\mu^2\right)\frac{\partial S}{\partial r_h}\;.
\end{equation} 
From $\partial S/\partial r_h \neq 0$ and since $\delta\mu \sim \epsilon$, this term above is $\sim 1/\epsilon$ unless 
\begin{equation}
\label{2ndmu}
\delta^2\mu =\frac{\delta\mu^2}{2\epsilon^2}
\end{equation}
Inserting \eq{2ndmu} back into \eq{muexp}, we obtain
\begin{equation}\label{WCCC_fin}
\mu = \epsilon^2 +\delta\mu\lambda+\frac{\lambda^2\delta\mu^2}{4\epsilon^2} =\left(\epsilon+\frac{\delta\mu\lambda}{2\epsilon}\right)^2\,.
\end{equation}  
This ensures that $\mu$ stays positive and WCCC holds. Due to lack of the explicit form of $S$, in the above proof we have only considered the marginal case of the second law, i.e., $\delta S=\delta^2 S=0$. However, in the explicit examples considered above, we do not need to require the regularity of $\delta^2 S$, so that we can consider the non-marginal cases, i.e., $\delta^2 S \ge 0$.



The notion of black hole entropy $S$ is well-defined only if the event horizon exists, i.e., $\mu>0$. 
{\rd This is not assuming what we want to show, as can be understood from the following perspective: for sufficiently small perturbations of a non-extremal black hole, the solution will certainly have a horizon; we can calculate the change in the entropy to the second order in this regime, and use this to show that at this order in perturbation theory $\mu$ is positive.} Moreover, the second law should be manifested from the underlying dynamical theory. The validity of our proposal implies that WCCC is guaranteed dynamically.  The nontrivial part of the proof is that the variation $\delta \mu$ due to the infalling matter is ${\cal O}(\epsilon)$ but the initial deviation from the extremality bound is ${\cal O}(\epsilon^2)$. It seems that the WCCC can be easily violated, but in fact it is not by requiring the second law. This is in the same spirit of the first-law approach by Sorce-Wald, in which the variation of entropy is assumed and used to evaluate the canonical energy.


\section{Discussion}
WCCC is important to protect a gravity theory from the pathology of naked singularity. In this work we propose and show that the second law of black hole thermodynamics ensures WCCC due to the peculiar dependence of the entropy on the extremality condition, and we explicitly demonstrate our proposal for a general class of quartic theories of gravity and electromagnetism. 

Naively, we expect to arrive the second law by the first law along with the energy condition of the infalling matter in Wald's gedanken experiment, {\rd however we find that this is not the case for our near-extremal charged black hole solutions in higher derivative gravity. In Appendix {\ref{sec:2}}  we show that the $n=1$ first law is apparently violated at ${\cal O}(c_i^2)$. This might be related to the gauge issue of Wald formalism. For gravity theory with fields with internal gauge freedom, one will expect the first law to be gauge invariant, however the chemical potential $\Phi_H$ depends explicitly on the gauge choice. This ambiguity may cause subtlety when applying the Sorce-Wald formalism straightforwardly to higher derivative gravity. The framework developed by Prabhu \cite{Prabhu:2015vua} using the principal bundle might be helpful to clarify this issue. On the other hand, since the entropy is gauge invariant, we can define the chemical potential as well as the Hawking temperature in terms of the variation of the entropy to derive a gauge invariant first law. This is just  what we have done in our second law approach. 

For general non-spherical collapsing case, the construction of the canonical energy would be quite involved in higher derivative gravity, and is crucial to check the second-order first law without source perturbation. Thorough treatment on this issue is expected for future study.
}

\acknowledgments
We thank Robert Wald for his helpful comments and discussions. {\rd We also thank the anonymous referee for the valuable insight.} FLL is supported by National Science and Technology Council (Taiwan), Grant 109-2112-M- 003-007-MY3. BN is supported by the National Natural Science Foundation of China with Grant No.~11975158 and 12247103. YC acknowledges the support from the Brinson Foundation, the Simons Foundation (Award Number 568762), and the National Science Foundation, Grants PHY-1708212 and PHY-1708213. 

\appendix
\widetext

\section{Checks of WCCC for quartic derivative theories of gravity and electromagnetism}\label{sec:1}
We show the explicit forms of the second-order perturbative solutions to the higher derivative theories with only one $c_i$ is turned on, as well as the corresponding details of checking WCCC via both the second law and the Sorce-Wald formalism.

\subsection{$c_1$ case}\label{App-c1}

The second-order solutions are solved by extending the procedure in \cite{Kats:2006xp} to $O(c_i^{~2})$. For the Lagrangian
\be
L \,=\, {1 \over 2\kappa} R - {1 \over 4} F_{\mu\nu}F^{\mu\nu} + c_1 R^2\,,
\ee
the solution is just the same as the one in Einstein-Maxwell theory, since the Ricci scalar $R$ of the unperturbed background is vanishing hence gives no contribution to the higher-order corrections of energy-momentum tensor. The check for  WCCC is also the same as in Einstein-Maxwell theory.

\subsection{$c_2$ case}\label{App-c2}

For the Lagrangian
\be
L \,=\, {1 \over 2\kappa} R - {1 \over 4} F_{\mu\nu}F^{\mu\nu} + c_2 R_{\mu\nu}R^{\mu\nu}\,,
\ee
the charged black hole solution turns out to be of the form
\be
ds^2 \,=\ -f(r) dt^2 + \frac{dr^2}{g(r) } + r^2 d\Omega  \label{a.met}
\ee
in which
\bea
f(r) &=& 1 - {\kappa m \over r} + {\kappa q^2 \over 2 r^2} 
\,+\, c_2 \left( - \,{2 \kappa^2 q^2 \over r^4 } + {\kappa^3 m q^2 \over r^5 } - {\kappa^3 q^4 \over 5 r^6} \right) \nn \\
&& ~ \,+ \, c_2^{~2} \left( { 48 \kappa^3 q^2 \over r^6 } - { 80 \kappa^4 m q^2 \over r^7 } 
+ {32 \kappa^5 m^2 q^2 \over r^8 } + {240 \kappa^4 q^4 \over 7 r^8 } 
- {51 \kappa^5 m q^4 \over 2 r^9} + {68 \kappa^5 q^6 \over 15 r^{10}} \right)\,,  \\
g(r) &=& 1 - {\kappa m \over r} + {\kappa q^2 \over 2 r^2} 
\,+\, c_2 \left( - \,{4 \kappa^2 q^2 \over r^4 } + {3 \kappa^3 m q^2 \over r^5 } - {6 \kappa^3 q^4 \over 5 r^6} \right) \nn \\
&& ~ \,+ \, c_2^{~2} \left( { 144 \kappa^3 q^2 \over r^6 } - { 304 \kappa^4 m q^2 \over r^7 } 
+ {160 \kappa^5 m^2 q^2 \over r^8 } + {1192 \kappa^4 q^4 \over 7 r^8 } 
- {351 \kappa^5 m q^4 \over 2 r^9} + {704 \kappa^5 q^6 \over 15 r^{10}} \right) 
\eea
with the gauge potential 
\be
A_t ~=~  - {q \over r} - c_2 {\kappa^2 q^3 \over 5 r^5} 
+ c_2^{~2} \left( {48 \kappa^3 q^3 \over 7 r^7 } - {8 \kappa^4 m q^3 \over r^8 } + {9 \kappa^4 q^5 \over 2 r^9} \right)\,.
\ee

For simplicity we set $\kappa = 2$ in the following. The existence of double root for either $f(r)=0$ or $g(r)=0$ determines the extremal condition 
\be \label{a.extc2}
m \,=\, |q| \left( 1 -  {4 c_2 \over 5 q^2 } -  {8 c_2^{~2} \over 21 q^4} \right)\,,
\ee 
or 
\be
|q| \,=\, m \left( 1 + {4 c_2 \over 5 m^2} -  {136 c_2^{~2} \over 525 m^4}  \right)\,.
\ee
The location of the horizon is also modified compared to the black hole solution in Einstein-Maxwell theory, 
\bea
r_h &=& r_0 \,+\,  \frac{4 c_2 q^2 (q^2 - 5 m r_0  + 5 r_0^{\,2}  )}{5 r_0^3 (m r_0 - q^2)} \nn \\
&&  -\;\, \frac{ 8 c_2^{~2} q^2 }{525 r_0^{\,7} (m r_0 - q^2)^3} 
 \Big( 4571 q^8 + 5 q^6 r_0 (3306 r_0 - 6881 m)  
 + 4200 m^2 r_0^{\,4} (3 r_0^{\,2} - 10 m r_0 + 8m^2)  \nn \\
&&   -\;\, 75 m q^2 r_0^{\,3}  (294 r_0^{\,2} - 1262 m r_0+ 1197 m^2)
 +  5 q^4 r_0^{\,2} (1995 r_0^{\,2} -14004 m r_0 + 17269 m^2) \Big)\,,
\eea
in which $\,r_0 \,=\, m + \sqrt{m^2 - q^2}$\;. The Hawking temperature could be obtained by the vanishing of the 
conical singularity for the corresponding Euclidean black hole,
\bea
T_H &=&  \frac{m r_0 - q^2}{2 \pi r_0^{\,3}} 
\,-\, \frac{2 c_2 q^2 (3 q^2 - 4 m r_0) ( 6 q^2 - 10 m r_0 + 5 r_0^{\,2} )}{5 \pi r_0^{\,7} (m r_0 - q^2)} \nn \\
&& +\;\, \frac{4 c_2^{~2} q^2}{525 \pi r_0^{\,11}(m r_0 - q^2)^3}
\Big( 245840 q^{10} - 4 q^8 r_0 (392462 m - 84765 r_0)   \nn \\
&& - \;\, 8400 m^3 r_0^{\,5} (88 m^2 - 74 m r_0 + 15 r_0^{\,2})
+ 150 m^2 q^2 r_0^{\,4} (20104 m^2 - 14216 m r_0 + 2275 r_0^{\,2})  \nn \\
&& - \;\, 50 m q^4 r_0^{\,3} (97916 m^2 - 54966 m r_0 + 6195 r_0^{\,2})
+ q^6 r_0^{\,2} (3943072 m^2 - 1575720 m r_0 + 93975 r_0^{\,2})  \Big) \,.
\eea

To obtain the Wald entropy, we first recall the Wald's formula \cite{Wald:1993nt,Iyer:1994ys}
\be
S \,=\, -\, 2\pi A \frac{\delta L}{\delta R_{\mu\nu\rho\sigma}} \epsilon_{\mu\nu} \epsilon_{\rho\sigma} {\Big|}_{r_h}
\ee
in which $A = 4 \pi r_h$ is the area of the horizon. For convenience we will introduce the null coordinate, i.e., define 
$ dv = \sqrt{f/g} \,dt + dr/g$\,, the metric (\ref{a.met}) then becomes
\be
ds^2  \,=\, 2 dv dr - g(r) dv^2 + r^2 d\Omega \,,
\ee
and the gauge potential ${\tilde A}_{\mu}$ in the null coordinates are 
\be
{\tilde A}_v = \sqrt{g \over f} \,A_t\,,  \qquad  {\tilde A}_r = -\, \sqrt{1 \over fg} \,A_t\,.
\ee
The Wald's formula then straightforwardly gives rise to 
\be
S \,=\, -\, 2 \pi A_h \left( -\, {1 \over \kappa}  - 4 c_2 R^{r v} \right) \,,
\ee
which turns out to be
\bea
S  &=& 4 \pi^2 r_0^{\,2} + \frac{ 32 c_2 \pi^2 q^2 (6 q^2 - 10 m r_0 + 5 r_0^2 )}{ 5 r_0^{2}(m r_0 - q^2)} \nn \\
&&  - \;\, \frac{64 c_2^{~2} \pi^2 q^2 }{525  r_0^{\,6}  (m r_0 - q^2 )^3}
  \Big( 41972 q^8  + 32 q^6 r_0 (1245 r_0 - 6503 m) 
  + 4200 m^2 r_0^{\,4}  (3 r_0^{\,2} - 16 m r_0 + 23 m^2) \nn \\
 && - \;\, 75 m q^2 r_0^{\,3}  (301 r_0^{\,2} - 2256 m r_0 + 4200 m^2) 
 + 20 q^4 r_0^{\,2} (525 r_0^{\,2} - 7134 m r_0 + 19243 m^2)   \Big) \,.
\eea

For near-extremal black holes, we introduce a small parameter $\epsilon$ to characterize the solution in such a way
\be \label{a.epsc2}
|q| \,=\, \sqrt{1 - \epsilon^2} \; m \left( 1 + {4 c_2 \over 5 m^2} -  {136 c_2^{~2} \over 525 m^4}  \right)\,.
\ee
Assuming that the first order variation is optimally done, i.e., the second law is satisfied marginally
\be
\delta S = 0 \,,
\ee
we obtain the following relation
\be \label{a.relc2}
\delta m \,=\, \delta q \,\Big( 1 - \epsilon + {\epsilon^2 \over 2 } 
+ \frac{2 c_2}{5 m^2} \left( 2 + 4 \epsilon - 27 \epsilon^2 \right) 
- \frac{ 4 c_2^{~2} }{175 m^4} \left( 6 + 8936 \epsilon - 88661 \epsilon^2 \right) \Big)\,.
\ee
The second order variation which satisfies the second law
\be
\delta^2 S \,\ge\, 0
\ee
gives rise to the inequality 
\be \label{a.ineqc2}
\delta^2 m \,\ge\, {1 \over m} (\delta q)^2 + \delta^2 q 
- \frac{4 c_2 }{5 m^3} \left(8 (\delta q)^2 - m \delta^2 q \right)
+ \frac{ 8 c_2^{~2} }{525 m^5} \left(27416 (\delta q)^2 - 9 m \delta^2 q \right) \,,
\ee
in which we have plugged in the relation (\ref{a.relc2})\,. From (\ref{a.extc2}) we know the WCCC is hold if 
\be \label{a.cric2}
W(m,q) \,\equiv\, m^2 - q^2  \left( 1 -  {4 c_2 \over 5 q^2 } -  {8 c_2^{~2} \over 21 q^4} \right)^2 \,\ge\, 0 \;.
\ee
To check (\ref{a.cric2}), consider a one-parameter family of solutions with $ m = m(\lambda)\,, \,q = q(\lambda) $\,. 
Expanding $W(m(\lambda), q(\lambda))$ to $O(\lambda^2)$ and using (\ref{a.epsc2})(\ref{a.relc2})(\ref{a.ineqc2}), 
we finally get 
\be
W(\lambda) \,\ge \, (\epsilon m - \lambda \delta q)^2  
+ \frac{8 c_2}{5 m^2} (\epsilon m - \lambda \delta q) (\epsilon m + 3 \lambda \delta q)
+ \frac{ 64 c_2^{~2} }{525 m^4} \left( 2 \epsilon^2 m^2 - 3351 \epsilon \lambda m \delta q 
+ 3433 \lambda^2 (\delta q)^2 \right)\;,
\ee
which could be recast to a perfect square 
\be
W(\lambda) \,\ge \, \Bigg[ \, \epsilon \left(  m + {4 c_2 \over 5 m} - {104 c_2^{~2} \over 525 m^3} \right)
- \lambda  \left(  1 - {12 c_2 \over 5 m^2 }  + {108344 c_2^{~2} \over 525 m^4 }\right) \delta q \, \Bigg]^2 
\,+\, O(c_2^{~3}) \,,
\ee
hence $W(\lambda) \,\ge \,0$ and WCCC is preserved up to $O(c_2^{~2})$ by the second law. 

On the other hand, according to the Sorce-Wald formalism \cite{Sorce:2017dst}, the first order variation is optimally done when 
\be
\delta m \;=\; \Phi_h \,\delta q \,,
\ee
from which we obtain the following relation
\be \label{a.relSWc2}
\delta m \,=\, \delta q \,\Big( 1 - \epsilon + {\epsilon^2 \over 2 } 
+ \frac{2 c_2}{5 m^2} \left( 2 + 4 \epsilon - 27 \epsilon^2 \right) 
- \frac{ 4 c_2^{~2} }{175 m^4} \left( 6 + 536 \epsilon - 4661 \epsilon^2 \right) \Big)\,.
\ee
which is slightly different from (\ref{a.relc2}) at $O(c_2^{~2})$. The second order variation inequality, 
\be
\delta^2 m - \Phi_h \delta^2 q \,\ge\, -\, \frac{T_H}{4 \pi} \delta^2 S^* \,,
\ee
combined with (\ref{a.relSWc2}) gives raise to 
\be \label{a.ineqSWc2}
\delta^2 m \,\ge\, {1 \over m} (\delta q)^2 + \delta^2 q 
- \frac{4 c_2 }{5 m^3} \left(8 (\delta q)^2 - m \delta^2 q \right)
- \frac{ 8 c_2^{~2} }{525 m^5} \left(10384 (\delta q)^2 + 9 m \delta^2 q \right) \,,
\ee
which is also different from (\ref{a.ineqc2}). Then $W(\lambda)$ turns out to satisfy
\be
W(\lambda) \,\ge \, (\epsilon m - \lambda \delta q)^2  
+ \frac{8 c_2}{5 m^2} (\epsilon m - \lambda \delta q) (\epsilon m + 3 \lambda \delta q)
+ \frac{ 64 c_2^{~2} }{525 m^4} \left( 2 \epsilon^2 m^2 - 201 \epsilon \lambda m \delta q 
-1292 \lambda^2 (\delta q)^2 \right)\;. 
\ee
The above expression could not be rewritten as a perfect square up to $O(c_2^{~2})$, as could be checked 
by examine the discriminant of the coefficients of $\lambda$ as in a quadratic equation. 
The best we can arrive is 
\be
W(\lambda) \,\ge \, \Bigg[ \, \epsilon \left(  m + {4 c_2 \over 5 m} + {50296 c_2^{~2} \over 525 m^3} \right)
- \lambda  \left(  1 - {12 c_2 \over 5 m^2 }  - {42856 c_2^{~2} \over 525 m^4 }\right) \delta q \, \Bigg]^2 
\,-\, {192  \epsilon^2 c_2^{~2} \over m^2 } \;,
\ee
hence WCCC is not guaranteed by the Sorce-Wald formalism.

\subsection{$c_3$ case}\label{App-c3}

For the Lagrangian
\be
L \,=\, {1 \over 2\kappa} R - {1 \over 4} F_{\mu\nu}F^{\mu\nu} + c_3 R_{\mu\nu\rho\sigma}R^{\mu\nu\rho\sigma}\,,
\ee
the solution is 
\bea
f(r) &=& 1 - {\kappa m \over r} + {\kappa q^2 \over 2 r^2} 
\,+\, c_3 \left( - \,{8 \kappa^2 q^2 \over r^4 } + { 4 \kappa^3 m q^2 \over r^5 } - { 4 \kappa^3 q^4 \over 5 r^6} \right) \nn \\
&& ~ \,+ \, c_3^{~2} \left( { 768 \kappa^3 q^2 \over r^6 } - { 1280 \kappa^4 m q^2 \over r^7 } 
+ { 512 \kappa^5 m^2 q^2  \over  r^8 } + {3840 \kappa^4 q^4 \over 7 r^8 } 
- {408 \kappa^5 m q^4 \over  r^9} + {1088 \kappa^5 q^6 \over 15 r^{10}} \right)\,,  \\
g(r) &=& 1 - {\kappa m \over r} + {\kappa q^2 \over 2 r^2} 
\,+\, c_3 \left( - \,{16 \kappa^2 q^2 \over r^4 } + {12 \kappa^3 m q^2 \over r^5 } - {24 \kappa^3 q^4 \over 5 r^6} \right) \nn \\
&& ~ \,+ \, c_3^{~2} \left( { 2304 \kappa^3 q^2 \over r^6 } - { 4864 \kappa^4 m q^2 \over r^7 } 
+ {2560 \kappa^5 m^2 q^2 \over r^8 } + {19072 \kappa^4 q^4 \over 7 r^8 } 
- {2808 \kappa^5 m q^4 \over  r^9} + {11264 \kappa^5 q^6 \over 15 r^{10}} \right)  \,,  \\
A_t &=&  - {q \over r} - c_3 {4\kappa^2 q^3 \over 5 r^5} 
+ c_3^{~2} \left( {768 \kappa^3 q^3 \over 7 r^7 } - {128 \kappa^4 m q^3 \over r^8 } + {72 \kappa^4 q^5 \over  r^9} \right)\,. 
\eea
The check for WCCC condition is straightforward just like the previous case, hence we just give the final result. The second law again gives raise to a perfect square
\be
W(\lambda) \,\ge \, \Bigg[ \, \epsilon \left(  m + {16 c_3 \over 5 m} - {1664 c_3^{~2} \over 525 m^3} \right)
- \lambda  \left(  1 - {48 c_3 \over 5 m^2 }  + {927104 c_3^{~2} \over 525 m^4 }\right) \delta q \, \Bigg]^2 
\,+\, O(c_3^{~3}) \,,
\ee
while the Sorce-Wald formalism gives
\be
W(\lambda) \,\ge \, \Bigg[ \, \epsilon \left(  m + {16 c_3 \over 5 m} + {401536 c_3^{~2} \over 525 m^3} \right)
- \lambda  \left(  1 - {48 c_3 \over 5 m^2 }  - {282496 c_3^{~2} \over 525 m^4 }\right) \delta q \, \Bigg]^2 
\,-\, {1536  \epsilon^2 c_3^{~2} \over m^2 }  \,+\, O(c_3^{~3}) \;.
\ee

\subsection{$c_4$ case}\label{App-c4}

For the Lagrangian
\be
L \,=\, {1 \over 2\kappa} R - {1 \over 4} F_{\mu\nu}F^{\mu\nu} + c_4 \kappa R F_{\mu\nu} F^{\mu\nu}\,,
\ee
the solution is 
\bea
f(r) &=& 1 - {\kappa m \over r} + {\kappa q^2 \over 2 r^2} 
\,+\, c_4  \left(  \,{4 \kappa^2 q^2 \over r^4 } - { 6 \kappa^3 m q^2 \over r^5 } + { 4 \kappa^3 q^4 \over  r^6} \right) 
 \,+ \, c_4^{~2} \left( - {32 \kappa^4 q^4 \over 7 r^8 } 
- {6 \kappa^5 m q^4 \over  r^9} + {32 \kappa^5 q^6 \over 3 r^{10}} \right)\,,  \\
g(r) &=& 1 - {\kappa m \over r} + {\kappa q^2 \over 2 r^2} 
\,+\, c_4 \left( - \,{16 \kappa^2 q^2 \over r^4 } + {14 \kappa^3 m q^2 \over r^5 } - {6 \kappa^3 q^4 \over  r^6} \right)  
\,+ \, c_4^{~2} \left(  {1088 \kappa^4 q^4 \over 7 r^8 } 
- {126 \kappa^5 m q^4 \over  r^9} + {152 \kappa^5 q^6 \over 3 r^{10}} \right) \,,   \\
A_t &=&  - {q \over r} - c_4 {2\kappa^2 q^3 \over  r^5} 
+ c_4^{~2} \left( {576 \kappa^3 q^3 \over 7 r^7 } - {96 \kappa^4 m q^3 \over r^8 } + {50 \kappa^4 q^5 \over  r^9} \right)\,. 
\eea
The second law gives raise to a perfect square for the WCCC condition
\be
W(\lambda) \,\ge \, \Bigg[ \, \epsilon \left(  m - {256 c_4^{~2} \over 21 m^3} \right)
- \lambda  \left(  1  + {211072 c_4^{~2} \over 21 m^4 }\right) \delta q \, \Bigg]^2 
\,+\, O(c_4^{~3}) \,,
\ee
while the Sorce-Wald formalism gives
\be
W(\lambda) \,\ge \, \Bigg[ \, \epsilon \left(  m + {161024 c_4^{~2} \over 21 m^3} \right)
- \lambda  \left(  1 -  {165248 c_4^{~2} \over 21 m^4 }\right) \delta q \, \Bigg]^2 
\,-\, {15360  \epsilon^2 c_4^{~2} \over m^2 } \,+\, O(c_4^{~3})  \;.
\ee

\subsection{$c_5$ case}\label{App-c5}

For the Lagrangian
\be
L \,=\, {1 \over 2\kappa} R - {1 \over 4} F_{\mu\nu}F^{\mu\nu} + c_5 \kappa R_{\mu\nu} F^{\mu\rho} F^{\nu}_{~\;\rho}\,,
\ee
the solution is 
\bea
f(r) &=& 1 - {\kappa m \over r} + {\kappa q^2 \over 2 r^2} 
\,+\, c_5  \left(  - {  \kappa^3 m q^2 \over r^5 } + { 4 \kappa^3 q^4 \over 5 r^6} \right) 
 \,+ \, c_5^{~2} \left( - {12 \kappa^4 q^4 \over 7 r^8 } 
+ {9 \kappa^5 m q^4 \over 2 r^9} - {164 \kappa^5 q^6 \over 45 r^{10}} \right)\,,  \\
g(r) &=& 1 - {\kappa m \over r} + {\kappa q^2 \over 2 r^2} 
\,+\, c_5 \left( - \,{6 \kappa^2 q^2 \over r^4 } + {5 \kappa^3 m q^2 \over r^5 } - {11 \kappa^3 q^4 \over 5 r^6} \right)  
\,+ \, c_5^{~2} \left(  {548 \kappa^4 q^4 \over 7 r^8 } 
- {139 \kappa^5 m q^4 \over 2 r^9} + {284 \kappa^5 q^6 \over 9 r^{10}} \right)  \,,  \\
A_t &=&  - {q \over r} - c_5 {\kappa^2 q^3 \over 5 r^5} 
+ c_5^{~2} \left( {48 \kappa^3 q^3 \over 7 r^7 } - {8 \kappa^4 m q^3 \over r^8 } + {43 \kappa^4 q^5 \over 6 r^9} \right)\,. 
\eea
The second law gives raise to a perfect square
\be
W(\lambda) \,\ge \, \Bigg[ \, \epsilon \left(  m + {4 c_5 \over 5 m} - {11512 c_5^{~2} \over 1575 m^3} \right)
- \lambda  \left(  1 - {12 c_5 \over 5 m^2 }  + {2469832 c_5^{~2} \over 1575 m^4 }\right) \delta q \, \Bigg]^2 
\,+\, O(c_5^{~3}) \,,
\ee
while the Sorce-Wald formalism gives
\be
W(\lambda) \,\ge \, \Bigg[ \, \epsilon \left(  m + {4 c_5 \over 5 m} + {1802888 c_5^{~2} \over 1575 m^3} \right)
- \lambda  \left(  1 - {12 c_5 \over 5 m^2 }  - {1763768 c_5^{~2} \over 1575 m^4 }\right) \delta q \, \Bigg]^2 
\,-\, {2304  \epsilon^2 c_5^{~2} \over m^2 }  \,+\, O(c_5^{~3}) \;.
\ee

\subsection{$c_6$ case}\label{App-c6}

For the Lagrangian
\be
L \,=\, {1 \over 2\kappa} R - {1 \over 4} F_{\mu\nu}F^{\mu\nu} + c_6 \kappa R_{\mu\nu\rho\sigma} F^{\mu\rho} F^{\rho\sigma}\,,
\ee
the solution is 
\bea
f(r) &=& 1 - {\kappa m \over r} + {\kappa q^2 \over 2 r^2} 
\,+\, c_6  \left(  - { 2 \kappa^2 q^2 \over r^4 } + { \kappa^3 m q^2 \over r^5 } - {  \kappa^3 q^4 \over 5 r^6} \right)  \nn\\
&&~ \,+ \, c_6^{~2} \left(  - {320 \kappa^4 m q^2 \over 7 r^7 } + {128 \kappa^5 m^2 q^2 \over 7 r^8 } 
 + {530 \kappa^4 q^4 \over 7 r^8 } 
- {411 \kappa^5 m q^4 \over 14 r^9} - {47 \kappa^5 q^6 \over 15 r^{10}} \right)\,,  \\
g(r) &=& 1 - {\kappa m \over r} + {\kappa q^2 \over 2 r^2} 
\,+\, c_6 \left( - \,{8 \kappa^2 q^2 \over r^4 } + {7 \kappa^3 m q^2 \over r^5 } - {16 \kappa^3 q^4 \over 5 r^6} \right)   \nn\\
&&~ \,+ \, c_6^{~2} \left( - {320 \kappa^4 m q^2 \over  r^7 } + {2048 \kappa^5 m^2 q^2 \over 7 r^8 } 
+ {4352 \kappa^4  q^4 \over 7 r^8 } 
- {1413 \kappa^5 m q^4 \over 2 r^9} + {3976 \kappa^5 q^6 \over 15 r^{10}} \right) \,,   \\
A_t &=&  - {q \over r} + c_6 \left( - { 2 \kappa^2 m q \over  r^4} + { 9\kappa^2 q^3 \over 5 r^5}  \right)
+ c_6^{~2} \left( - {64 \kappa^4 m^2 q \over 7 r^7 } - {160 \kappa^3 q^3 \over 7 r^7 } 
+ {216 \kappa^4 m q^3 \over 7 r^8 } - {9 \kappa^4 q^5 \over 10 r^9} \right)\,. 
\eea
The second law gives raise to a perfect square
\be
W(\lambda) \,\ge \, \Bigg[ \, \epsilon \left(  m + {4 c_6 \over 5 m} - {10504 c_6^{~2} \over 525 m^3} \right)
- \lambda  \left(  1 - {52 c_6 \over 5 m^2 }  + {1850344 c_6^{~2} \over 525 m^4 }\right) \delta q \, \Bigg]^2 
\,+\, O(c_6^{~3}) \,,
\ee
while the Sorce-Wald formalism gives
\be
W(\lambda) \,\ge \, \Bigg[ \, \epsilon \left(  m + {4 c_6 \over 5 m} + {1199096 c_6^{~2} \over 525 m^3} \right)
- \lambda  \left(  1 - {52 c_6 \over 5 m^2 }  - {972056 c_6^{~2} \over 525 m^4 }\right) \delta q \, \Bigg]^2 
\,-\, {4608  \epsilon^2 c_6^{~2} \over m^2 }  \,+\, O(c_6^{~3}) \;.
\ee

\subsection{$c_7$ case}\label{App-c7}

For the Lagrangian
\be
L \,=\, {1 \over 2\kappa} R - {1 \over 4} F_{\mu\nu}F^{\mu\nu} 
+ c_7 \kappa^2 F_{\mu\nu} F^{\mu\nu} F_{\rho\sigma}F^{\rho\sigma}\,,
\ee
the solution is 
\bea
f(r) \,=\, g(r) &=& 1 - {\kappa m \over r} + {\kappa q^2 \over 2 r^2} - { 4 c_7  \kappa^3 q^4 \over 5 r^6}
+ { 128 c_7^{~2}  \kappa^5 q^6 \over 9 r^{10}} \,, \\
A_t &=&  - {q \over r} + { 16 c_7 \kappa^2 q^3 \over 5 r^5}  - {256 c_7^{~2}   \kappa^4 q^5 \over 3 r^9} \,.
\eea
In this case, the second law and the Sorce-Wald formalism give raise to the same perfect square
\be
W(\lambda) \,\ge \, \Bigg[ \, \epsilon \left(  m + {16 c_7 \over 5 m} - {48256 c_7^{~2} \over 225 m^3} \right)
- \lambda  \left(  1 - {48 c_7 \over 5 m^2 }  + {236416 c_7^{~2} \over 225 m^4 }\right) \delta q \, \Bigg]^2 
\,+\, O(c_7^{~3}) \,.
\ee

\subsection{$c_8$ case}\label{App-c8}

For the Lagrangian
\be
L \,=\, {1 \over 2\kappa} R - {1 \over 4} F_{\mu\nu}F^{\mu\nu} 
+ c_8 \kappa^2 F_{\mu\nu} F^{\nu\rho} F_{\rho\sigma}F^{\sigma\mu}\,,
\ee
the solution is 
\bea
f(r) \,=\, g(r) &=& 1 - {\kappa m \over r} + {\kappa q^2 \over 2 r^2} - { 2 c_8  \kappa^3 q^4 \over 5 r^6}
+ { 32 c_8^{~2}  \kappa^5 q^6 \over 9 r^{10}} \,, \\
A_t &=&  - {q \over r} + { 8 c_8 \kappa^2 q^3 \over 5 r^5}  - {64 c_8^{~2}   \kappa^4 q^5 \over 3 r^9} \,.
\eea
In this case, the second law and the Sorce-Wald formalism again give raise to the same perfect square
\be
W(\lambda) \,\ge \, \Bigg[ \, \epsilon \left(  m + {8 c_8 \over 5 m} - {12064 c_8^{~2} \over 225 m^3} \right)
- \lambda  \left(  1 - {24 c_8 \over 5 m^2 }  + {59104 c_8^{~2} \over 225 m^4 }\right) \delta q \, \Bigg]^2 
\,+\, O(c_8^{~3}) \,.
\ee

\subsection{$c_2 + c_4$ case}\label{App-c2c4}
We also consider the case with both $c_2$ and $c_4$ terms are turned on, i.e.,
\be
L \,=\, {1 \over 2\kappa} R - {1 \over 4} F_{\mu\nu}F^{\mu\nu} 
+ c_2 R_{\mu\nu}R^{\mu\nu} + c_4 \kappa R F_{\mu\nu} F^{\mu\nu}\,,
\ee
the solution for which is a bit tedious, hence we just show the final result for WCCC. The second law still gives a perfect square
 \bea
  W(\lambda) &\ge& \Bigg[ \, \epsilon \left(  m + {4 c_2 \over 5 m} 
- {8 \times ( 13 c_2^{~2} + 800 c_4^{~2} ) \over 525 m^3} \right) \nn \\
&&~  - \lambda  \left(  1 - {12 c_2 \over 5 m^2 }  + { 8 \times (13543 c_2^{~2} + 193200 c_2 c_4 + 659600 c_4^{~2} ) \over 525 m^4 }\right) \delta q \, \Bigg]^2  \,+\, O(c_i^{~3})  \,, 
\eea
while the Sorce-Wald formalism gives
 \bea
  W(\lambda) &\ge& \Bigg[ \, \epsilon \left(  m + {4 c_2 \over 5 m} 
- {8 \times ( 6287 c_2^{~2} + 113400 c_2 c_4 + 503200  c_4^{~2} ) \over 525 m^3} \right)  \nn \\
&&~ - \lambda  \left(  1 - {12 c_2 \over 5 m^2 }  + { 8 \times (5357 c_2^{~2} + 113400 c_2 c_4 + 516400 c_4^{~2} ) \over 525 m^4 }\right) \delta q \, \Bigg]^2  \nn \\
&&~  - \, { 192  \epsilon^2  ( c_2^{~2} + 18 c_2 c_4 + 80 c_4^{~2} ) \over m^2 } \,+\, O(c_i^{~3})  \,.
\eea

\section{Check of first law up to {\cal O}($c_i$)} \label{sec:2}
 
The standard first law of black hole thermodynamics without source perturbation is to set $\delta^n q_{\rm H}=\delta^n T_{ab}=0$ in eq. (16) of the main text. That is, 
\be
\delta^n m_{\rm ADM} -\Phi_{\rm H} \delta^n q_B - T_H \delta^n S_B 
 = \delta_{n,2} {\cal E}_{\Sigma}(\phi; \delta \phi, {\cal L}_{\xi}\phi)
\ee
Since we do not have the explicit form of the canonical energy ${\cal E}_{\Sigma}(\phi; \delta \phi, {\cal L}_{\xi}\phi)$, we will check only the $n=1$ first law for the black hole solutions considered in this work. As argued in the main text, the higher derivative corrections will not affect the relation $\delta^ m_{\rm ADM}=\delta^n m$ due to the higher powers of $1/r$ suppression at infinity. For $\delta^n q_B=\delta^n (\int_B \epsilon_{abcd} S^{cd})$ with $S^{ab}=F^{ab}+{\cal O}(c_i)$, we can either evaluate $q_B$ at the bifurcation sphere $B$, i.e., $q_B=\sqrt{g^{tt}g^{rr}} r^2 S_{rt}|_{r=r_+}$, or at spatial infinity by adopting Gauss theorem and the no-source assumption, it is then straightforward to see $\delta^n q_B=\delta^n q$. Therefore, the $n=1$ first law implies
\be
\Delta_{\rm I}:= {\partial S_B \over \partial q} + \Phi_{\rm H} {\partial S_B \over \partial m} =0\;. 
\ee
For all cases of black hole solutions up to ${\cal O}(c_i^2)$ considered in the main text, we find that 
\be 
\Delta_{\rm I} = 0 + {\cal O}(c_i^2)\;.
\ee 
The ${\cal O}(c_i^2)$ terms are finite in extremal limit. For example, for the $c_2$ case,
\be
\Delta^{(c_2)}_{\rm I}= {1536\pi^2 (1-8\epsilon + {\cal O}(\epsilon^2)) \over M^3} c_2^2 + {\cal O}(c_i^3)\;
\ee
where $\epsilon$ is the small non-extremility parameter.

From the above, we can conclude that all the black hole solutions considered in this work satisfy the $n=1$ first law up to ${\cal O}(c_i)$. This fact is crucial in \cite{Chen:2020hjm} to prove the WCCC for the extremal black holes of generic gravity theories by the Sorce-Wald formalism. The ${\cal O}(c_i^2)$ violation for the $n=1$ first law {\rd should be responsible for the failure of the Sorce-Wald formalism to yield the WCCC for near-extremal black holes as discussed in the main text.}
The check of $n\ge 2$ sourceless first law for modified gravity theories is beyond the scope of this work because it needs the explicit construction of the canonical energy.

\section{Implication on WCCC from Afloat Spherical Thin-Shell} \label{sec:3}   

Since we are considering the spherical collapsing to avoid the complications due to the electromagnetic and gravitational radiations, the simplest example is the spherical collapsing shell. To find the implication on WCCC condition, we will consider the spherical thin-shell matter afloat in the spacetime described by the metric \eq{a.met} with $f$ and $g$  given in  section \ref{sec:1} up to ${\cal O}(c_i)$. 

   The motion of the thin shell around a black hole obeys the generalized Israel junction conditions \cite{Israel:1966rt}, which can be obtained from the Gauss-Codazzi equations. However, the junction conditions for thin-shell are in general highly singular in the higher derivative gravity theories except for the Gauss-Bonnet higher derivative term, see for example \cite{Davis:2002gn,Deruelle:2000ge,Balcerzak:2007da,Chu:2021uec} for discussions. To have the regular junction conditions to yield sensible motion of thin-shell, we need to impose regularity conditions on the metric around the thin-shell. For more singular junction conditions, it means more regularity conditions on the metric should be imposed so that mostly it will yield only trivial solutions, i.e., no thin-shell. See section \ref{3rd-order} for some discussion. Below we will only consider the thin-shell in Einstein and Einstein-Gauss-Bonnet gravities. 
   
   The junction conditions for Einstein gravity are given by \cite{Israel:1966rt} (set $\kappa=2$)
\be\label{JC-Einstein}
\Big[K_{\mu\nu}-h_{\mu\nu} K\Big]_J=-S_{\mu\nu}
\ee
where $K_{\mu\nu}$ is the extrinsic curvature, $K\equiv K^{\mu}_{\nu}$, $h_{\mu\nu}$ is the induced metric on the thin-shell, and $S_ab$ is the stress tensor of the thin-shell matter. Here $\Big[A\Big]_J$ denotes taking difference of the quantity $A$ on the both sides of the thin-shell. Assume the spherical thin-shell is located at $r=r_s$ with stress tensor $S^{\mu}_{\nu}=\textrm{diag}(\rho,0,p,p)$, and evaluate the extrinsic curvature with the metric \eq{a.met}, the junction conditions give
\be
\Big[g\Big]_J=-{r_s\over 2} \rho\;.
\ee
and
\be\label{JC_EM_2}
\Big[\sqrt{g} (2+ r f'/f)\Big]_J= 2 r_s p\;.
\ee
where $f'\equiv \partial_r f$. It turns out that the above junction conditions involve only $g$, $f$ and $f'$. To have a finite jump on the left sides of the above junction conditions, we only need to impose the piecewise continuity of $f$ at $r=r_s$ to 
to yield sensible and nontrivial junction conditions. For this purpose, we choose to rescale the coordinate time so that the metrics on both sides of the thin-shell are given by
\be
f_+(r)=g_+(r)=1-2{m_+ \over r}+ {q_+^2 \over r^2} \;,  
\ee
but
\be
{1-2{m_- \over r_s}+ {q_-^2 \over r_s^2} \over 1-2{m_+ \over r_s}+ {q_+^2 \over r_s^2}} f_-(r)=g_-(r)=1-2{m_- \over r}+ {q_-^2 \over r^2}  
\ee
Note that $f_+(r_s)=f_-(r_s)$.  We assume the matter shell is pressure-less, i.e., $p=0$, then the junction condition \eq{JC_EM_2} can be turned into the following condition
\be\label{thin-shell-cons}
m_+^2 -q_+^2 =\Big({ r_s - m_+ \over r_s -m_- }\Big)^2 (m_-^2 -q_-^2)\;.
\ee
This condition implies that a sub-extremal black hole with $m_-^2> q_-^2$ remains sub-extremal, i.e., $m_+^2> q_+^2$ even after throwing a pressure-less spherical thin-shell. This is consistent with WCCC. 


   Next we will show that the same condition also holds for the Einstein-Gauss-Bonnet (EGB) gravity. Note that for EGB gravity, we shall introduce the coupling $c_{EGB}$ of the Gauss-Bonnet term, which is nothing but $c_{EGB}=c_1=c_2=-{1\over 4}c_2$. The junction condition for EGB gravity are different from the one for Einstein gravity and are given by \cite{Davis:2002gn}
\be\label{JC_EGB}
\Big[ K_{\mu\nu}-h_{\mu\nu} K +2 c_{EGB} (3 J_{\mu\nu}-h_{\mu\nu} J + 2 \hat{P}_{\mu\rho\lambda\nu} K^{\rho\lambda})\Big]_J=-S_{\mu\nu}
\ee
where 
\be
J_{\mu\nu}={1\over 3}(2 K K_{\mu\rho}K^{\rho}_{\nu} +K_{\rho\lambda}K^{\rho\lambda}K_{\mu\nu}-2K_{\mu\rho}K^{\rho\lambda}K_{\lambda\nu}-K^2 K_{\mu\nu})\;,
\ee 
$J\equiv J^{\mu}_{\nu}$ and 
\be
\hat{P}_{\mu\nu\rho\lambda}=\hat{R}_{\mu\nu\rho\lambda}+2\hat{R}_{\nu[\rho}h_{\lambda]\mu}-2\hat{R}_{\mu[\rho}h_{\lambda]\nu} + \hat{R}_{\nu[\rho}h_{\lambda]\mu}+ h_{\mu[\rho}h_{\lambda]\nu} \hat{R} 
\ee 
where the hatted quantities are the associated unhatted quantities evaluated with respect to the induced metric $h_{\mu\nu}$. The novelty of the junction condition \eq{JC_EGB} is only the first derivatives of the metric are involved. Using the metric \eq{a.met} and the induced metric for the spherical thin-shell, we can find that $\hat{P}_{\mu\nu\rho\lambda}=0$ and $3 J_{\mu\nu}-h_{\mu\nu} J=0$ even though $J_{\mu\nu}$ and $J$ are nonzero. Based on the above, the junction condition \eq{JC_EGB} for EGB gravity is indeed reduced to \eq{JC-Einstein} for Einstein gravity. Moreover, the Gauss-Bonnet term is a total derivative term so that it will not affect the field equation, and the black hole solutions are the same as the one for Einstein-Maxwell theory. In total, the junction condition of the EGB gravity will still yield the same constraint \eq{thin-shell-cons} for the spherical thin-shell. That is, the thin-shell will not turn a sub-extremal black hole into a naked singularity in the EGB gravity. This is consistent with our result for EGB gravity as discussed in the main text.  

\subsection{No thin-shell from a third-order junction condition} \label{3rd-order}

In \cite{Balcerzak:2007da} a set of third-order junction conditions for the higher derivative gravity theories have been proposed. This junction condition is obtained by collecting the singular terms in the Gauss-Codazzi equations. For the quartic action of gravity considered in the main text, the junction conditions take the following form
\be\label{W3_JC}
\Big[W^{\mu}_{\nu\rho}\Big]_J n^{\rho} = 2 S^{\mu}_{\nu}
\ee
where $n^{\mu}$ is the normal vector of the thin-shell, and 
\bea
W^{\mu}_{\nu\rho}&=&c_1 R_{;\lambda} (2 \delta^{\mu}_{\nu} \delta^{\lambda}_{\rho} -g^{\mu\lambda} g_{\nu\rho} -\delta^{\lambda}_{\nu} \delta^{\mu}_{\rho}) \nn \\
&+&c_2 R^{\mu}_{\nu;\rho} + {1\over 2} g^{\mu}_{\nu} \hat{R}_{;\rho} - R^{\mu}_{\rho;\nu}- R_{\rho\nu}^{;\mu})-4 c_3 R^{\;\mu\;\lambda}_{\rho\;\nu;\lambda}\;. 
\eea
After explicitly evaluating the left-side of \eq{W3_JC} with respect to the metric \eq{a.met}, the result involves the terms with $g''$ and $f'''$. This implies that we need to impose the continuity conditions $\Big[f\Big]_J=\Big[g\Big]_J=\Big[f'\Big]_J=\Big[g'\Big]_J=\Big[f''\Big]_J=0$ to yield a finite left-side of \eq{W3_JC}, thus a sensible junction condition with finite $S^{\mu\nu}$.   Since the junction condition \eq{W3_JC} is already ${\cal O}(c_i)$, thus the metric used to evaluate the junction condition should be kept up to the leading order only. Thus, the metrics on both sides of the thin-shell contain in total only four integration constants, i.e., $m_{\pm}$ and $q_{\pm}$.  The above five continuity conditions are over-constrained on these integration constants and can be shown only to yield trivial solutions, namely, $m_+=m_-$ and $q_+=q_-$. This implies no sensible thin-shell for generic quartic gravities,

\section{Dependence of $r_h$ on $m$ and $q_j$ near the extremal boundary} \label{sec:4}

  Suppose we have $m$ and $q_j$, and $m = m_{\rm ex}(q_j)$ is the mass of extremal black holes, with black holes given by $m\ge m_{\rm ex}(q_j)$.   Let us define $\mu = m - m_{\rm ex}(q_j)$.  Let us also denote $r_h(\mu,q_j)$ the horizon radius, with $R(q_j) = r_h(0,q_j)$ the radius of extremal black holes as a function of $q_j$.  Let us first show that for $\mu$ inside an open neighborhood of 0, 
\begin{equation}
\label{eqrexp}
r_h(\mu,q_j ) = R(q_j) + \sqrt{\mu} \, {\rd \rho} (q_j,\sqrt{\mu})
\end{equation}
with ${\rd \rho}$ a smooth function of its arguments.  In GR, the horizon radius is the greater real root of the quadratic polynomial  $r_h^2 - 2 M r_h +Q^2=0$,
\begin{equation}
r_\pm = M\pm \sqrt{M^2-Q^2}
\end{equation}
 with extremal boundary given by the location of the double root. 

 In modified GR, we can view the $(r_h, {\rd \mu, q_j})$ relation in two ways.  We can directly write a condition of 
\begin{equation}
\label{eq:Delta}
\Delta (r_h ,\mu , q_j)=0
\end{equation}
 with extremal condition given by  
 \begin{equation}
\partial_r\Delta |_{R(q_j),0,q_j}=0
\end{equation}
As we expand around $\mu =0$, and write $r_h$ as an expansion, 
\begin{equation}
r_h = R(q_j) + \delta r_h\,,
\end{equation}
we have
\begin{align}
0&=\Delta (R(q_j)+\delta r_h,\mu,q_j) \nonumber\\
&=
\frac{1}{2}\frac{\partial^2 \Delta}{\partial r^2}\Bigg|_ {R(q_j),0,q_j} (\delta r_h)^2 +
\frac{\partial \Delta}{\partial \mu}\Bigg|_ {R(q_j),0,q_j}\mu +\ldots
  \end{align}
 From this, we can re-expand $\,{\rd \delta r_h} \,$ in terms of $\sqrt{\mu}$:
 \begin{equation}
 \delta r_h =\sqrt{\mu} \sum_{n=0}^{+\infty}  \alpha_n(q_j)  \mu^{n/2}  =\sqrt{\mu} \, {\rd \rho}(q_j,\sqrt{\mu})\,.
 \end{equation}   
 
 Another way to write this, is to view Eq.~\eqref{eq:Delta} as a definition of $\mu$ in terms of $r_h$ and $q_j$.  In the GR case, we have
\begin{equation}
M(r_h,Q)=\frac{r_h^2+Q^2}{2r_h}
\end{equation}
For each $Q$, we generically have two values of $r_h$ that gives rise to $M$ --- only the larger value correspond to the outer horizon.  Extremal black holes are when $M$ takes a minimum at $r_h = Q$.  In the modified gravity case, expecting the structure of the problem to remain unchanged, namely the fact that ${\rd m}$  is uniquely determined by $r_h$ and ${\rd q_j}$ and the fact that ${\rd m}$ is a minimum when $r_h(q_j) = R(q_j)$.  We need to assume that when deviating away from the minimum, the value of ${\rd m}$ depends quadratically on $ r-R{\rd (q_j)}$\,, 
\begin{equation}
\mu (r,q_j) ={\rd m}(r,q_j)-{\rd m}(R(q_j),q_j) =[r-R(q_j)]^2 F(r,q_j)
\end{equation}
with $F(r,q_j)$ a smooth, non-zero function in an open  neighborhood of $(R(q_j),q_j)$. We can then write
\begin{equation}
r =R(q_j) +\frac{\sqrt{\mu}}{\sqrt{F(r,q_j)}}
 \end{equation}
 This can be solved iteratively to yield Eq.~\eqref{eqrexp}.

\end{document}